\documentstyle[psfig,epsf]{article}
\def\beq{\begin{equation}}
\def\eeq{\end{equation}}
\def\bea{\begin{eqnarray}}
\def\eea{\end{eqnarray}}
\begin{document}

\begin{center}
{\large \bf Phase diagram and critical exponents of a Potts gauge
        glass}\\[.3in] 
        {\bf Jesper Lykke Jacobsen (1) and Marco
Picco (2)} \\  
	{\bf (1)} {\it Laboratoire de Physique Th\'eorique et Mod\`eles
                   Statistiques, Universit\'e Paris-Sud, \\
                   B\^atiment 100, 91405 Orsay, France.}\\
	{\bf (2)} {\it LPTHE\/},\\
        {\it  Universit{\'e} Pierre et Marie Curie et 
              Universit{\'e} Denis Diderot\\
	      Bo\^{\i}te 126, Tour 16, 1$^{\it er}$ {\'e}tage \\
	      4 place Jussieu,
	      F-75252 Paris Cedex 05, France}\\

\end{center}

\begin{abstract}
The two-dimensional $q$-state Potts model is subjected to a $Z_q$ symmetric
disorder that allows for the existence of a Nishimori line.  At $q=2$, this
model coincides with the $\pm J$ random-bond Ising model.  For $q>2$, apart
from the usual pure and zero-temperature fixed points, the
ferro/paramagnetic phase boundary is controlled by {\em two} critical fixed
points: a weak disorder point, whose universality class is that of the {\em
ferromagnetic} bond-disordered Potts model, and a strong disorder point
which generalizes the usual Nishimori point. We numerically study the case
$q=3$, tracing out the phase diagram and precisely determining the critical
exponents. The universality class of the Nishimori point is inconsistent
with percolation on Potts clusters.
\end{abstract}

During the last decade, the study of disordered systems has attracted much
interest. This is true in particular in two dimensions, where the possible
types of critical behavior for the corresponding pure models can be
classified using conformal field theory \cite{BPZ}. Recently, similar
classification issues for disordered models have been addressed through the
study of various random matrix ensembles \cite{Z}, but many fundamental
questions remain open.

An important category of 2D disordered systems is given by models where the
disorder couples to the local energy density. Two paradigmatic members of
this class are the $\pm J$ random-bond Ising model, and the $q$-state
ferromagnetic random-bond Potts model. The model to be studied in the
present Letter can be thought of as an interpolation between these two
members; we shall therefore begin by recalling some of their basic
properties.

The random-bond Ising model (RBIM) is defined by the energy functional
\beq
  {\cal H}_{\rm Ising} =
  \sum_{\langle i,j \rangle} J_{ij} \delta(S_i,S_j)  \, ,
  \label{RBI}
\eeq
where the sum is over the edges of the square lattice, $S_i = \pm 1$
are Ising spins, and $\delta(.,.)$ is the Kronecker delta
function. The random bonds take the values $J_{ij}=\pm 1$ according to
the probability distribution
\beq
  P(J_{ij}) = p \delta(J_{ij}-1) + (1-p) \delta(J_{ij}+1) \, .
 \label{pmJprob}
\eeq
The salient feature of this model is that it marries disorder with
frustration, leading to the possibility of spin glass order.

\begin{figure}
\centerline{\psfig{file=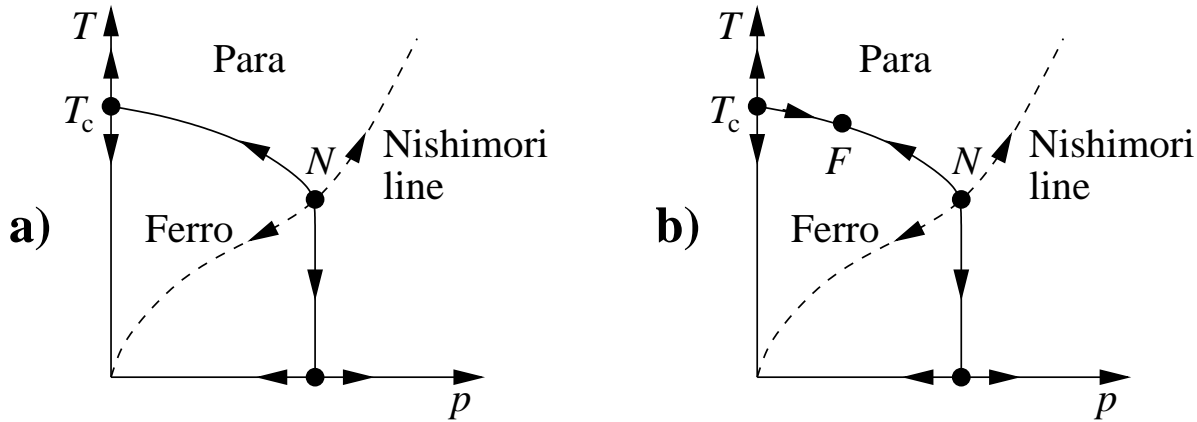,width=1.0\columnwidth}}
\caption{
Phase diagram of the $\pm J$ random-bond Ising model (a)
and the $q>2$ state Potts gauge glass (b).
\label{phase}
}
\end{figure}

Its phase diagram is generally believed to be as in Fig.~\ref{phase}.a
\cite{LG}.
The boundary FP between the ferromagnetic and the paramagnetic phases
is controlled by three fixed points. The attractive fixed points at
either end of the phase boundary are respectively the critical point
of the pure Ising model and a zero-temperature fixed point.
Between these two we find the multicritical point N, intersecting the
so-called Nishimori line \cite{N}
\beq
  {\rm e}^{\beta} =  (1-p)/p \, .
\label{NishimoriLine}
\eeq
On this line, the replicated version of the model possesses a local
$Z_2$ gauge symmetry that, among other things, allows for exactly
computing the internal energy and for establishing the pairwise
equality of correlation functions
\beq
\label{NCF}
 [ \langle S_{i_1}S_{i_2}\cdots S_{i_k} \rangle^{2n-1} ] =
 [ \langle S_{i_1}S_{i_2}\cdots S_{i_k} \rangle^{2n} ] \, , 
\eeq
where $\langle \cdots \rangle$ denotes the thermal and $[ \cdots ]$
the disorder average.  Since the Nishimori line is also invariant
under Renormalization Group (RG) transformations \cite{LG}, its
intersection N with the FP boundary must be a fixed point.  However,
the widespread belief that the corresponding universality class is of
the percolation type has recently been refuted on the basis of
numerical evidence \cite{HPP}.

The other model of special interest to us is the random-bond Potts
model (RBPM), which is also defined by (\ref{RBI}), except that the
spins now take $q$ different values, $S_i=1,2,\ldots,q$. The most
well-studied case is that of purely ferromagnetic bonds, such as
\beq
  P(J_{ij}) = \frac12 \delta(J_{ij}+J_1) + \frac12 \delta(J_{ij}+J_2),
\eeq
with $R \equiv J_2/J_1 \ge 1$ adjusting the disorder strength.

In contradistinction to the Nishimori point, the fixed point of this model
is situated at weak disorder. For $q>2$ the disorder is relevant
\cite{Harris}, and the corresponding line of fixed points tends to the one
of the pure Ising model in the limit $q\to 2$. As a consequence, the
critical exponents can be computed perturbatively in a
$(q-2)$-expansion \cite{Perturb}. According to the RG picture, for
$q>2$ any small amount of disorder should induce a flow towards the
random fixed point. That this is also true for $q>4$, where the phase
transition in the pure model is of the first order, is the content of
the Aizenman-Wehr theorem \cite{AW}.

In this Letter we shall consider the model
\beq
 {\cal H} = -\sum_{\langle i,j \rangle} \delta^{(q)}(S_i-S_j+J_{ij})  \, ,
 \label{PGG}
\eeq
where $S_i=1,2,\ldots,q$, and $\delta^{(q)}(x)=1$ if $x=0 \mbox{ mod }
q$ and zero otherwise. The randomness now takes the form of a local
``twist'' $J_{ij}$, which is clearly a more severe type of disorder
than simple bond randomness. The variables $J_{ij}$ are taken from the
distribution
\beq
 P(J_{ij}) = \big(1-(q-1)p \big) \delta(J_{ij}) +
              p \sum_{J=1}^{q-1} \delta(J_{ij}-J) \, ,
\eeq
with $0 \le p \le 1/(q-1)$ controlling the strength of the randomness. We
shall refer to this model, which was originally introduced in
Ref.~\cite{ns}, as 
the Potts Gauge Glass (PGG). The particular form of the randomness ensures
the existence of a Nishimori line (see below).  For $q=2$, the PGG reduces
to the RBIM, and for $p=1/q$ it was studied analytically in \cite{Gold}.
It is also connected to the RBPM: To wit, when $q>2$ the pure Potts model
($p=0$) should be {\em unstable} to a small amount of randomness, meaning
that the RG flow cannot be as indicated on Fig.~\ref{phase}.a. Instead, we
are forced to assume the existence of a new fixed point F, intermediary
between the pure model and the Nishimori point (see Fig.~\ref{phase}.b).
But whenever $(q-2)$, and hence the value of $p$ at F, is sufficiently
small, frustration effects are negligible, and we should flow to the {\em
same} random fixed point as in the RBPM. For reasons of continuity we
expect this argument to hold true also for higher values of $q$.

The expression of the Nishimori line was obtained in Ref.~\cite{ns}, but
since our notation is slightly different we shall repeat the argument
here. We first reexpress  
the disorder distribution as
\beq
P(J_{ij}) = p e^{K \delta^{(q)}(J_{ij})} \; \; \hbox{with} \; \;
K=\log \big( 1/p-(q-1) \big).
\label{newdist}
\eeq
Consider then the disorder averaged internal energy 
\beq
 E = {\cal N} \sum_{\{J_{ij}\}} \left( \prod_{\langle ij \rangle}
 e^{K \delta^{(q)}(J_{ij})} \right) \times \frac{\sum_{\{S_i\}}
 \delta^{(q)}(S_i-S_j+J_{ij}) e^{-\beta {\cal H}}} 
         {\sum_{\{S_i\}} e^{-\beta {\cal H}}},
\eeq
where ${\cal N} = -1/(q-1+e^K)^{2N}$, $N$ being the number of sites of
the square lattice. ${\cal H}$ is then invariant under the gauge
transformation $S_i \rightarrow S_i - \sigma_i, J_{ij} \rightarrow
J_{ij}+\sigma_i-\sigma_j$, though $P(J_{ij})$ is not.  Still, $E$ is
invariant since we sum over all configurations of the disorder. Then,
averaging over all the possible gauge transformations we get
\beq
 E = {\cal N}q^{-N} \sum_{\{J_{ij}\}} \sum_{\{\sigma_i\}}
 \left( \prod_{\langle ij \rangle} e^{K \delta^{(q)}(J_{ij}+\sigma_i-\sigma_j)}
 \right) \times  \frac{\sum_{\{S_i\}} \delta^{(q)}(S_i-S_j+J_{ij})
 e^{-\beta {\cal H}}}  {\sum_{\{S_i\}} \prod_{\langle i'j' \rangle}
         e^{\beta \delta^{(q)}(S_{i'}-S_{j'}+J_{i'j'})}}.
\eeq
Imposing $K=\beta$, there is a remarkable simplification:
\bea
 E &=& {\cal N} q^{-N} \sum_{\{J_{ij}\}} \sum_{\{S_i\}}
       \delta^{(q)}(S_i-S_j+J_{ij}) e^{-\beta {\cal H}} \nonumber \\
   &=& {\cal N} q^{-N} \frac{\partial}{\partial \beta}
       \sum_{\{S_i\}} (e^{\beta}+q-1)^{2N} = 
       \frac{-2N e^\beta}{q-1+e^\beta}.
\eea
Thus $E$ is regular, and Eq.~(\ref{newdist}) with $K=\beta$ defines
the generalized Nishimori line.

Normalized two-point functions are defined by
\beq
\label{eq12}
 \langle S_i S_j \rangle = (q-1)^{-1} 
 \big( q \langle \delta^{(q)}(S_i-S_j) \rangle - 1 \big).
\eeq
Let us now recall how Eq.~(\ref{NCF}) can be derived for $q=2$. We
consider $n=1$ for simplicity.  Using the trivial identities
$\delta^{(q)}(\Delta S- \Delta \sigma) =
 \sum_{l=0}^{q-1} \delta^{(q)}(\Delta S-l) \delta^{(q)}(\Delta \sigma+l)$
and
$\sum_{l=0}^{q-1} \delta^{(q)}(\Delta S-l)=1$
one readily establishes that 
\beq
\label{fact}
 2 \delta^{(2)}(\Delta S-\Delta \sigma)-1 =
\big( 2 \delta^{(2)}(\Delta S) -1 \big) \big (2
\delta^{(2)}(\Delta \sigma) -1 \big),
\eeq
and, using the same gauge transformation as before,
\beq
[ 2 \langle \delta^{(2)}(S_i-S_j) \rangle -1 ] = 
[ (2 \langle \delta^{(2)}(S_i-S_j) \rangle -1)^2] .
\eeq
This relies crucially on the fact that the above trivial identities
generate only two terms, and for general $q$ we do not expect simple
relations like (\ref{NCF})\footnote{A simple relation for a chiral-type
correlator was established in Ref.~\cite{ns}, but in terms of
Eq.~(\ref{eq12}) this does not lead to degeneracy in the multiscaling
spectrum.}. 

We now turn to our numerical results.  Random transfer matrices in the
Fortuin-Kasteleyn (FK) representation \cite{FK} have been a very
powerful tool for studying the RBPM \cite{JC}.  Unfortunately, the
random twist variables $J_{ij}$ present in (\ref{PGG}) complicate the
definition of the FK clusters: only those clusters are allowed for
which $\sum_\gamma J_{ij} = 0 \mbox{ mod } q$ for any path $\gamma$
within the cluster \cite{CGN}. It is not obvious how this constraint
can be generalized to {\em real} values of $q$, and even for integer
$q$ keeping track of the necessary local information would greatly
increase the number of basis states needed.  We have therefore found
it more convenient to write the transfer matrices directly in the spin
basis.  We work at $q=3$ throughout, but expect our conclusions to
extend to arbitrary $q>2$.

As we have shown in an earlier publication \cite{JP}, the phase
diagram can be traced out by investigating the effective central
charge.  To this end we have computed the free energy $f_L^{(p)} =
{\ln Z^{(p)} \over L M}$ on strips of various widths $L$ and
practically infinite length, $M=10^5$.  The (effective) central charge
$c$ can then be obtained as the universal coefficient of the
finite-size correction to the free energy for periodic boundary
conditions \cite{central}
\beq
 f_L^{(p)} = f_{\infty}^{(p)} + {c \pi \over 6 L^2} + \cdots.
 \label{cDef}
\eeq

According to Zamolodchikov's $c$-theorem \cite{Zamc}, here applied to
a non-unitary theory, the effective central charge {\em increases}
along the RG flows and coincides with the (true) central charge at the
fixed points.  The FP boundary (cf.~Fig.~\ref{phase}) can be
traced by identifying the maximum of $c$ as a function of $T$, for
various fixed values of $p$.

Since the randomness is strong, and since the fits to (\ref{cDef})
must be based on at least two different sizes $L$ to eliminate the
non-universal quantity $f_{\infty}^{(p)}$, we have taken several
precautions in order to obtain small error bars on the $f_L^{(p)}$.
First, for any fixed value of $p$ we use the {\em same} realization of
the disorder for the computations at different values of $T$.  Second,
for each strip of length $M=10^5$ we work in a canonical ensemble,
meaning that disorder realizations for which the fraction of bonds
$J_{ij} = J$ does not {\em exactly} equal $p$ for each
$J=1,2,\ldots,q-1$ are discarded. Third, for each strip we average
$f_L^{(p)}$ over up to $10^5$ independent realizations.

\begin{table}
\begin{center}
\begin{tabular}{l|llllllll}
$p$ & $L=3,4$  &     & $L=4,5$ &     & $L=2,3,4$ &     & $L=3,4,5$ & \\
    & $\beta$  & $c$ & $\beta$ & $c$ & $\beta$   & $c$ & $\beta$   & $c$ \\
\hline
 0.01 
      &  1.0521(5) & 0.76825(3) & 1.0520(5) & 0.78074(9)
      &  1.0520(5) & 0.79460(6) & 1.0520(5) & 0.7987(3) \\
 0.02 
      &  1.1061(5) & 0.76874(6) & 1.1061(5) & 0.7815(2)
      &  1.1061(5) & 0.7953(1)  & 1.1061(5) & 0.7998(6) \\
 0.03 
      &  1.1692(5) & 0.76907(9) & 1.1691(5) & 0.7816(2)
      &  1.1692(5) & 0.7957(2)  & 1.1691(5) & 0.7997(6) \\
 0.04 
      &  1.244(1)  & 0.7685(1)  & 1.245(1)  & 0.7822(7)
      &  1.244(1)  & 0.7951(3)  & 1.245(1)  & 0.8020(17) \\
 0.05 
      &  1.336(1)  & 0.7663(3)  & 1.337(1)  & 0.7799(9)
      &  1.336(1)  & 0.7925(6)  & 1.338(1)  & 0.7995(25) \\
 0.06 
      &  1.453(2)  & 0.7620(3)  & 1.456(2)  & 0.7739(11)
      &  1.454(2)  & 0.7882(6)  & 1.456(2)  & 0.7911(30) \\
\end{tabular}
\end{center}
\caption{\label{tabcc}Parametrisation of the ferro/paramagnetic phase
boundary.} 
\end{table}

In Table~\ref{tabcc} we show the resulting values of $c$ and the
inverse temperature $\beta=1/T$ at the FP boundary. The two-point fits
are based directly on (\ref{cDef}), while the three-point fits include
an additional non-universal $1/L^4$ correction \cite{JC}. The
existence of an attractive fixed point at $p \sim 0.04$ with a central
charge slightly larger than $c_{\rm pure} = 4/5$, characterizing the
pure 3-state Potts model, is brought out very clearly.

The reader may wonder why data for such small system sizes can
possible give any reliable information about the thermodynamic limit.
Comparison with the pure model ($p=0$) shows however that in
particular the three-point fits converge very rapidly towards the
exact result: $c_{3,4} = 0.76803$, $c_{4,5} = 0.78043$, $c_{2,3,4} =
0.79431$, $c_{3,4,5} = 0.79831$ \cite{note}.  We have extrapolated the
data at the fixed point F by assuming that for each fit, the relative
deviation from the infinite-size result is the same as in the pure
model. In this way we arrive at the final result
\beq
 c_{\rm F} = 0.8025(10),
\eeq
which compares favorably with the perturbative result $c_{\rm pert} =
\frac{4013}{5000} + {\cal O}(q-2)^5 \approx 0.8026$
\cite{Perturb} for the {\em ferromagnetic} RBPM.

To numerically locate the Nishimori point we measure $c_{\rm eff}$
along the Nishimori line. Since in this case $p$ is a function of
$\beta$ (see Eq.~(\ref{newdist}) with $K=\beta$) we can no longer work
in the canonical ensemble of disorder realizations. Accordingly our
error bars are larger.  It is however a big advantage to know the
exact parametrisation of the Nishimori line, since otherwise we would
have had to scan a two-dimensional manifold of parameter values
\cite{Sorensen}. 

\begin{table}
\begin{center}
\begin{tabular}{l|llllllll}
 $p$   & $L=3,4$   & $L=4,5$    & $L=2,3,4$  & $L=3,4,5$ \\
\hline
 0.077 
       & 0.7208(4) & 0.7284(22) & 0.7374(8)  & 0.739(6) \\
 0.078 
       & 0.7212(5) & 0.7346(27) & 0.7374(10) & 0.754(7) \\
 0.079 
       & 0.7218(5) & 0.7316(22) & 0.7386(11) & 0.746(6) \\
 0.080 
       & 0.7213(7) & 0.7292(24) & 0.7379(14) & 0.741(6) \\
\end{tabular}
\end{center}
\caption{\label{tabnishi}Effective central charge along the Nishimori
line.} 
\end{table}

From the data in Table~\ref{tabnishi} we conclude that the fixed point
N is located at $p_{\rm N} = 0.0785(10)$. Using the same extrapolation
procedure as above we also estimate
\beq
 c_{\rm N} = 0.756(5).
\eeq
This is in remarkable agreement with the value of the central charge
for the percolation limit in the RBPM: $c= \frac{5 \sqrt{3} \ln{q}}{4
\pi} \approx 0.7571$ \cite{JC}.  Below we shall return to the question
whether the Nishimori point is ``just'' percolation.

We have also measured magnetic multiscaling exponents $\eta_k$,
defined in the plane by
\beq
[\langle S(x_1,y_1) S(x_2, y_2)\rangle^{2k}]= \big( (x_2-x_1)^2 +
(y_2-y_1)^2 \big)^{-\eta_k/2}  
\label{2by2}
\eeq
for any integer $k$. On the semi-infinite cylinder of circumference
$L$, with $x\in [1,L]$ and $y\in ]-\infty,+\infty[$, this reads, using
a conformal mapping,
\beq
[\langle S(x_1,y) S(x_2,y)\rangle ^n] \propto
  \left(\sin\left({\pi (x_2-x_1)\over L}\right) L\right)^{- \eta_{n}}. 
\label{fitsscf}
\eeq
For a pure system, $\eta_n = n \times \eta$, while for percolation
over Potts clusters all $\eta_n$ coincide. The principal goal is here
to establish the non-trivial multiscaling at N, rather than to
determine the $\eta_k$ with extraordinary precision.  The largest
system size employed was $L=12$, and we approximate the semi-infinite
cylinder by taking a length of $M=400 L$. All runs were averaged over
$10^3$ disorder configurations.

\begin{figure}
\centerline{\psfig{file=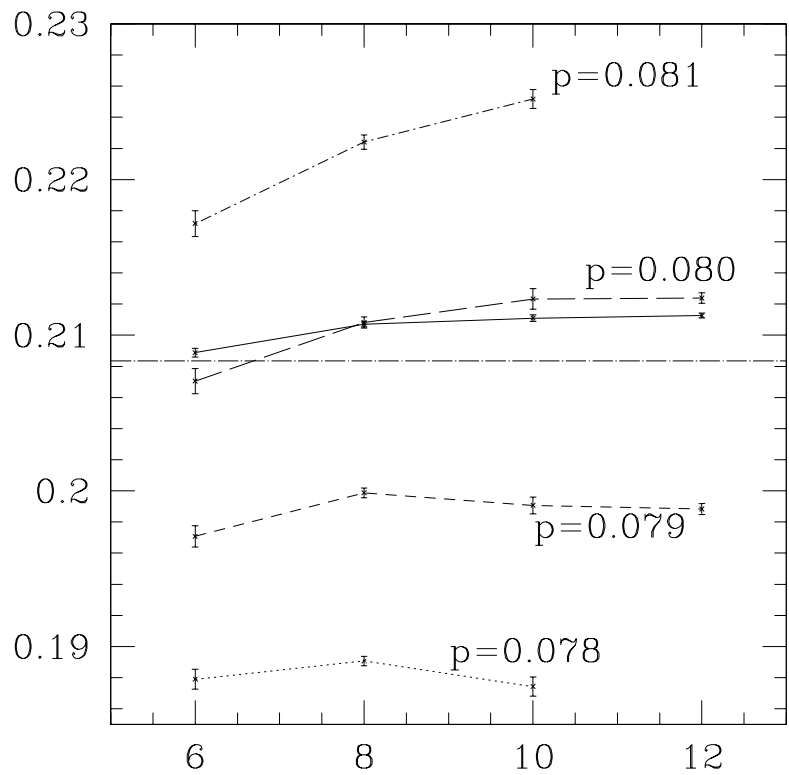,width=0.7\columnwidth}}
\caption{$\eta_1$ extracted from (\ref{fitsscf}) with
$\Delta x = 1,\ldots,L$. We also show the corresponding fit for
percolation (full line) and the exact value $\eta_{\rm perc} = 5/24$.
\label{fspin}
}
\end{figure}

In Fig.~\ref{fspin}, we show effective values of $\eta_1(L)$ along the
Nishimori line, for various $p$ close to $p_{\rm N}$.  These values
were obtained by fitting data for all $x_2-x_1=1,\ldots,L/2$ to
(\ref{fitsscf}); to judge the systematic error due to the inclusion of
the smallest $\Delta x \equiv |x_2-x_1|$ we also display a similar
plot for ordinary percolation, where $\eta_{\rm perc} = 5/24 \simeq
0.2083$ is known exactly.  At the fixed point, $\eta_1(L)$ must tend
to a constant, and we conclude that $p_{\rm N} = 0.079-0.080$ with
$\eta_1 = 0.20-0.21$.  Discarding the smallest $\Delta x$ leads to
consistent results, but with larger error bars.

Although our value of $\eta_1$ is consistent with percolation,
this scenario can be excluded by considering higher moments.
E.g.~for $p=0.080$ and $L=12$ we obtain
\bea
\eta_1=0.21239 (35) \; \; \; \; \eta_2=0.25192 (39) \nonumber \\ 
\eta_3=0.30824 (47) \; \; \; \; \eta_4= 0.33773 (52),
\eea
the corresponding values for $p=0.079$ being some 6 \% smaller.

\begin{figure}
\centerline{\psfig{file=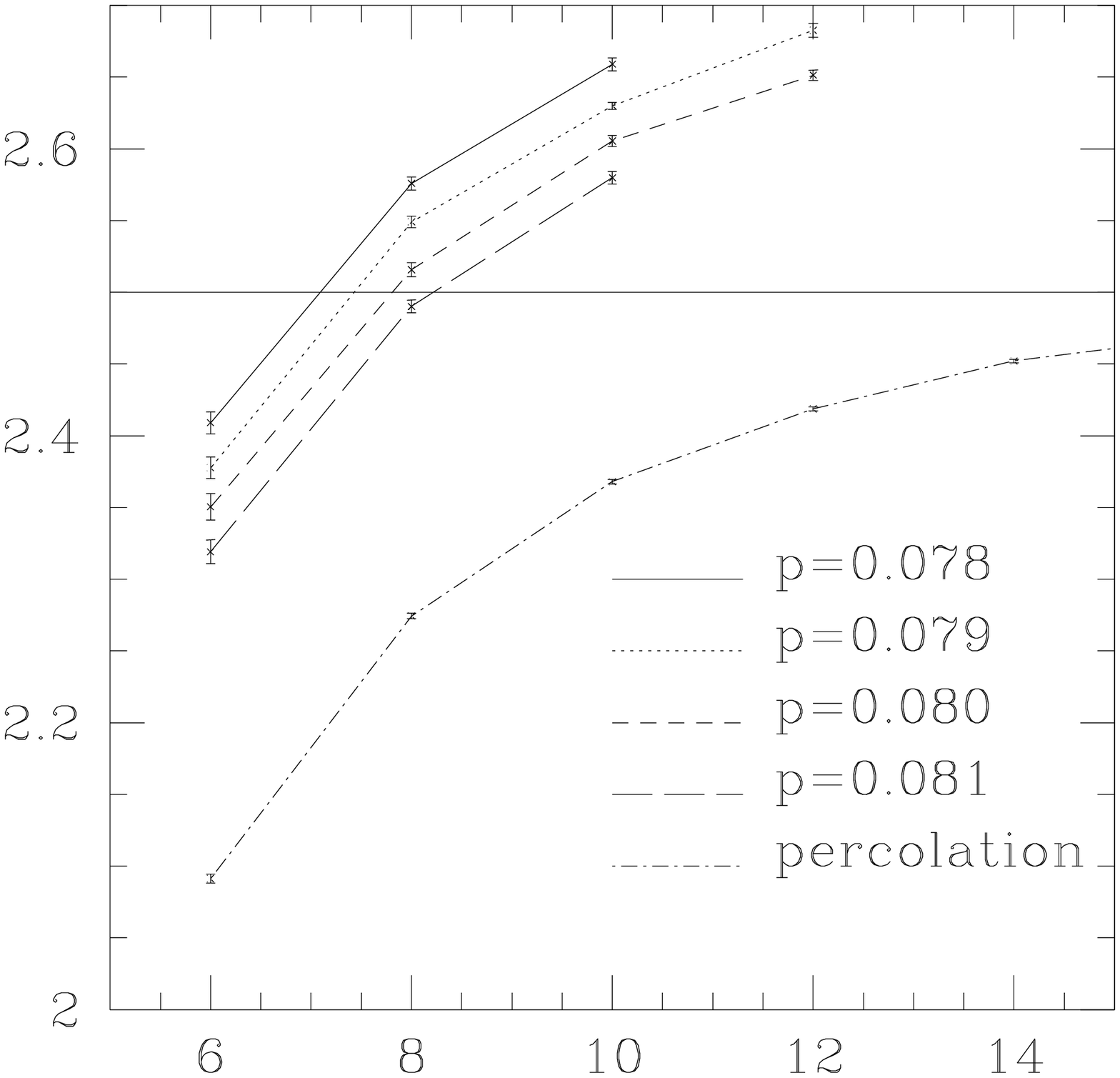,width=0.7\columnwidth}}
\caption{$\eta^{\rm e}_1$ extracted from (\ref{fitsscf}) with
$\Delta x = 1,\ldots,L$. We also show the corresponding fit for
percolation and the exact value $\eta^{\rm e}_{\rm perc} = 5/2$. 
\label{fen1}
}
\end{figure}

Further evidence against percolation can be obtained by similarly
considering the energy-energy correlations. In analogy with the RBIM
case we associate this with a deviation from N along the {\em
vertical} direction on Fig.~\ref{phase}.b. The results for $\eta^{\rm
e}_1$ are shown in Fig. \ref{fen1}, and once again we
compare with the percolation value $\eta^{\rm e}_{\rm perc}  =
2(2-1/{\nu_{\rm perc}})= 5/2$. 
In this case, the exponents depend less on the precise value of
$p_{\rm N}$, but the finite-size corrections are larger.
Extrapolating, we find a value of roughly $\eta^{\rm e}_1 =2.75-2.85$,
rather close to the one obtained for the RBIM Nishimori point
$\eta^{\rm e}_1=2.83(2)$ using a similar fit \cite{HPP2}.  Discarding data
with small $\Delta x$ leads to larger error bars, but is still
consistent with $\eta^{\rm e}_1 \sim 2.85$. We have also verified that 
the energy correlations exhibit genuine multiscaling.

In conclusion, we have studied a $q$-state (Potts-like)
generalization of the $\pm J$ random-bond Ising model that allows for
the definition of a Nishimori line.  Apart from a weak disorder fixed
point that coincides with that of the well-studied random-bond Potts
model, the model possesses a strong disorder point with multiscaling
exponents different from those of percolation. The fixed point
structure is reminiscent of that found by S\o rensen {\em et al.}
\cite{Sorensen} in the context of a $\pm J$ like Potts model, which
does however not possess the gauge symmetry required for defining a
Nishimori line. We believe that it would be interesting to study
whether the critical points of these two models are indeed identical.
Open questions concerning our model include the study of its
zero-temperature limit, the possibility of reentrance, and of its
behavior for $q>4$. It would also be interesting to examine it using a
supersymmetric approach.

We would like to thank J.~Cardy, A.~Honecker and P.~Pujol for useful
discussions.


\begin{thebibliography}{99}

\bibitem{BPZ}
A.~A.~Belavin, A.~M.~Polyakov and A.~B.~Zamolodchikov,
Nucl.~Phys.~B {\bf 241}, 333 (1984);
J.~Stat.~Phys.~{\bf 34}, 763 (1984).

\bibitem{Z}
M.~R.~Zirnbauer, J.~Math.~Phys.~{\bf 37}, 4986 (1996);
A.~Altland and M.~R.~Zirnbauer, Phys.~Rev.~B {\bf 55}, 1142 (1997).

\bibitem{LG}
W.~L.~McMillan, Phys.~Rev.~B~{\bf 29}, 4026 (1984); 
A.~Georges and P.~Le Doussal,
unpub\-lis\-hed pre\-print (1988); P.~Le Doussal and
A.~B.~Harris, Phys. Rev.~Lett.~{\bf 61}, 625 (1988),
Phys.~Rev.~B~{\bf 40}, 9249 (1989).  

\bibitem{N}
H.~Nishimori,
Prog.~Theor.~Phys.~{\bf 66}, 1169 (1981),
J.~Phys.~Soc.~Jpn.~{\bf 55}, 3305 (1986);
Y.~Ozeki and H.~Nishimori,
J.~Phys.~A~{\bf 26}, 3399 (1993).

\bibitem{HPP}
A.~Honecker, M.~Picco and P.~Pujol, to be published in Phys.~Rev.~Lett. and
cond-mat/0010143.

\bibitem{Harris}
A.~B.~Harris, J.~Phys.~C {\bf 7}, 1671 (1974).

\bibitem{Perturb}
A.~W.~W.~Ludwig and J.~L.~Cardy, Nucl.~Phys.~B {\bf 285}, 687 (1987);
A.~W.~W.~Ludwig, Nucl.~Phys.~B {\bf 330}, 639 (1990);
Vl.~Dotsenko, M.~Picco and P.~Pujol, Nucl.~Phys.~B {\bf 455}, 701 (1995).

\bibitem{AW}
M.~Aizenman and J.~Wehr, Phys.~Rev.~Lett.~{\bf 62}, 2503 (1989).

\bibitem{ns} H.~Nishimori and M.~J.~Stephen, Phys.~Rev.~B {\bf 27}, 5644
(1983). 

\bibitem{Gold}
Y.~Y.~Goldschmidt, J.~Phys.~A {\bf 22}, L157 (1989).

\bibitem{FK}
P.~W.~Kasteleyn and C.~M.~Fortuin,
J.~Phys.~Soc.~Jap.~{\bf 46} (suppl.), 11 (1969).

\bibitem{JC}
J.~L.~Jacobsen and J.~L.~Cardy, Nucl.~Phys.~B {\bf 515}, 701 (1998).

\bibitem{CGN}
M.~Caselle, F.~Gliozzi and S.~Necco, J.~Phys.~A.~{\bf 34}, 351 (2001).

\bibitem{JP}
J.~L.~Jacobsen and M.~Picco, Phys.~Rev.~E {\bf 61}, R13 (2000).

\bibitem{central}
H.~W.~J.~Bl\"ote, J.~L.~Cardy and M.~P.~Nightingale,
Phys.~Rev.~Lett.~{\bf 56}, 742 (1986);
I.~Affleck, {\em ibid.}~{\bf 56}, 746 (1986).

\bibitem{Zamc}
A.~B.~Zamolodchikov,
Pis'ma Zh.~Eksp.~Teor.~Fiz.~{\bf 43}, 565 (1986).
[JETP Lett.~{\bf 43}, 730 (1986).]

\bibitem{note}
We have chosen to transfer along the $(1,1)$ direction of the square lattice,
since (at least for the pure model) this yields a faster convergence and,
more importantly, a monitonic one.

\bibitem{Sorensen}
E.~S.~S\o rensen, M.~J.~P.~Gingras and D.~A.~Huse,
Euro.~Phys.~Lett.~{\bf 44}, 504 (1998).

\bibitem{HPP2}
A.~Honecker, M.~Picco and P.~Pujol, to be published.

\end{thebibliography}
\end{document}